\documentclass[12pt]{article}
\usepackage{graphicx,amssymb,epsfig,epsf} 
 \usepackage{ulem}
\usepackage[usenames]{color}
\usepackage{rotating}
\usepackage{epsfig}
\usepackage{amsmath}
\begin{document}

\begin{flushright}
   {\bf IPMU12-0145}\\
\end{flushright}

\vskip 30pt

\begin{center}
{\large \bf Degenerate SUSY search at the 8 TeV LHC}\\
\vskip 20pt
{Biplob Bhattacherjee$^{a}$\footnote{bbhattacherjee@gmail.com} and {Kirtiman Ghosh$^{b}$\footnote{kirti.gh@gmail.com}}}  \\
\vskip 10pt
{$^a$Kavli Institute for the Physics and Mathematics of the Universe (WPI),\\
The University of Tokyo, Kashiwa, Chiba 277-8583, Japan. }\\
\vskip 10pt
{ $^{b}$
Department of Physics and Helsinki Institute of Physics\\
FIN-00014, University of Helsinki, Finland.
 }

\end{center}

\vskip 20pt
\abstract{
ATLAS and CMS collaborations have searched for the supersymmetry (SUSY) 
in different channels with 7 TeV center-of-mass energy and their 
search strategies are optimized for the SUSY scenarios with fairly 
large mass splittings among the sparticles. In absence of any significant 
deviation of the data from the Standard Model (SM) prediction, stringent 
bounds are imposed on the squarks and gluino masses in the context of 
different SUSY scenarios. Since the exact SUSY breaking mechanism is not 
known, some part of the SUSY spectrum may be degenerate. In this paper, 
we consider two particular cases of degenerate SUSY spectrum and discuss 
the impact of ATLAS and CMS SUSY searches on these scenarios. We found 
that present SUSY search strategies of the ATLAS and CMS collaborations 
are not optimized for the degenerate SUSY scenarios. Even the LHC with 8 TeV 
center-of-mass energy can not probe large part of the parameter space 
of degenerate SUSY scenarios with the present search strategies. 
}

\section{Introduction}
One of the major goals of the Large Hadron Collider (LHC) experiment is to find new physics beyond the Standard Model (SM). Perhaps, supersymmetry (SUSY) \cite{SUSY} is the most attractive alternative beyond the SM due to the following reasons: (i) SUSY offers a solution to the hierarchy problem, (ii) it predicts  grand unification in gauge interactions and (iii) moreover, ${\cal R}$ parity conserving SUSY scenarios naturally provide a weakly interacting massive stable lightest SUSY particle (LSP) which can be a good candidate for the cold dark matter. Therefore, search for SUSY is of the prime targets of the LHC.

The LHC is a proton-proton collider and thus, the collision processes are overwhelmed by the QCD interactions. Therefore, most of the generic SUSY searches rely on the strong productions of squarks ($\tilde{q}$) and/or gluino ($\tilde{g}$). The decay of squark/gluino can be classified easily by assuming a specific type of mass hierarchy. For gluino heavier than squarks, $\tilde{g} \rightarrow \tilde{q} q$ is expected to be the dominant decay mode. In case of squarks heavier than gluino, three body decays of gluino $\tilde{g} \rightarrow q q \tilde{\chi}$ to electroweak gauginos ($\tilde{\chi}$) through virtual squarks are important. Similarly, squarks dominantly decay to quark(s) plus gauginos. Gauginos further decay to LSP in association with one or more SM particles. In ${\cal R}$ parity conserving (RPC) SUSY scenarios, LSP is stable and weakly interacting and thus, it remains invisible in the detector giving rise to missing transverse energy ($E_T\!\!\!\!\!/~$) signature. The characteristic signature for the pair production of squarks and/or gluinos (in RPC scenarios) is multijets plus $E_T\!\!\!\!\!/~$ in association with leptons. The number of jets can be roughly estimated from the decay modes described above. The presence of third generation squarks (or sleptons) may change the above conclusions and the decay chains containing third generation squarks (or sleptons) have to be treated separately. With 4.7 (4.89) fb$^{-1}$ integrated luminosity of proton-proton collision at 7 TeV center-of-mass energy,  no SUSY signature has been observed in the multijets plus $E_T\!\!\!\!\!/~$ channel by the ATLAS \cite{multijet_ATLAS} (CMS \cite{multijet_CMS}) collaboration. As a result, squark and gluino masses have been excluded below about 1 TeV at 95\% confidence level (C.L.) for the mSUGRA/cMSSM model \cite{cmssm}. However, it is important to mention that mSUGRA/cMSSM type of SUSY breaking scenario generally leads to the wide separation between gluino/squark and LSP mass. As a result, the decay products of gluino or squarks are generally hard and thus, the signatures can be easily separated from SM background by demanding hard jets or leptons in association with $E_T\!\!\!\!\!/~$ in the final state.

If SUSY is the working principle beyond the SM, the null results in the SUSY searched at the LHC indicate towards the following possibilities:
\begin{itemize}
\item The colored SUSY particles (squarks and gluinos) are very heavy and thus, their production cross-sections at the 7 TeV LHC are highly suppressed. In this case, the signature of electroweak SUSY sector can only show up at the LHC. However, we have to accumulate large amount of data to extract the signal from the background. 
\item  Another possibility is that squarks and/or gluinos are not heavy, but, they are nearly degenerate with the LSP. The hardness of the signal jets/leptons depends on the relative separation between squark/gluino and LSP, not on the mass of the produced particles. In case of nearly degenerate mass spectrum, jets or leptons arising from the decay of SUSY particles may be soft and it may even fall below the detector acceptance level. The pair production cross-section of squarks and/or gluinos will be high in this case, however, the signature may be buried deeply in the background events. Therefore, the discovery of this kind of scenarios can  be really challenging at the LHC \cite{comSUSY,Dreiner}.
\end{itemize}

The initial state radiation (ISR) that results from the quark/gluon leg of squark/gluino production process only depends on the scale of the interaction and color structure. However, the hardness of ISR jets do not depend on the relative separation between produced particles and decay products. The mass of the strongly interacting SUSY particles (squark and gluino) under consideration can be of the order of few hundreds of GeV to TeV. Thus, we expect that the ISR jets arises with the pair production of squark and/or gluino should be hard enough to be detected \cite{Alwall:2008va}. In presence of hard radiation, the pair of squark/gluino system will recoil against the ISR jets and thus, giving rise to  $E_T\!\!\!\!\!/~$ comparable with the transverse momentum of the jets. Therefore, if squarks and/or gluinos are nearly degenerate with the LSP then jets (coming from ISR) plus $E_T\!\!\!\!\!/~$ is the only signature of squark and/or gluino pair production. However, in such cases the jet multiplicity is smaller than the multiplicity expected in non-degenerate SUSY scenarios and thus, the Standard Model background could be severe. ATLAS and CMS collaborations demand very hard $E_T\!\!\!\!\!/~$ and effective mass cuts to extract SUSY signature from the SM background. These cuts are optimized for non-degenerate SUSY spectrum as in the case of cMSSM. However, the applicability of ATLAS and CMS SUSY search strategies for degenerate SUSY spectrum is poorly studied and understood. Therefore, it is important to study whether the present ATLAS or CMS SUSY search strategies are good enough for degenerate SUSY spectrum or not. 

Tagging of ISR jet is successfully applied for constraining dark matter models in which dark matter couples to SM quarks or gluons through higher dimensional operators. The pair production of dark matter particles with ISR jets is the only way to probe such particles at the collider experiments. The collider searches for the degenerate SUSY scenarios are somewhat similar to the above mentioned example. In principle, degenerate SUSY scenarios can be probed using ISR jets in monojet of multijets channels. However, the optimized channel and cuts for the degenerate SUSY search are still not known. 

In this article, we consider two particular cases of degenerate SUSY mass spectrum. For {\bf Scenario I}, we consider gluino to be nearly degenerate with the LSP and other sparticles are decoupled. This scenario is well motivated from WMAP \cite{wmap} relic density (RD) data. In most SUSY scenarios, the lightest neutralino is the LSP and it can be a viable dark matter candidate. However, for most combination of parameters the computed LSP relic density can not explain the WMAP data. The RD predicted for the SUSY scenarios is either too large (if the LSP is bino–like) or too small (if it is higgsino– or wino–like). However, the observed relic density can be explained if a bino-like neutralino co-annihilate with a nearly degenerate gluino. {\bf Scenario II} is similar to scenario {\bf Scenario I} apart from the fact that now the squarks are not too heavy to be produced at the LHC.  If squarks are not too heavy, squark-gluino pair production also contributes to the multijets $+E_T\!\!\!\!\!/~$ signature. Moreover, squarks decay to gluino gives rise to a hard jet at the parton level. Therefore, it is important to study the gluino mass reach in the context of nearly degenerate gluino-NLSP and neutralino-LSP scenario when the squarks are heavier that gluino however, within the reach of the LHC.

The paper is organized as follows. We define our model points in the next section. Section 3 contains summary of the ATLAS (CMS) multijets (monojets) $+E_T\!\!\!\!\!/~$ search strategies and detailed description about our signal and background analysis. In section 4, the bounds and possible discovery reach of SUSY particles are discussed. In Section 4, we also discuss about the applicability of jets $+E_T\!\!\!\!\!/~$ signature for other degenerate BSM scenarios like minimal Universal Extra Dimension model. Finally, we summarize our work in Section 5. 

\section{Model description}
SUSY searches at the collider experiments are really complicated. The complicacy arises from the unknown SUSY breaking mechanism which makes the 
sparticle spectrum completely arbitrary. As for example, in order to specify masses and couplings of minimal $\mathcal{R}$ parity conserving scenario (MSSM), we require more than 100 parameters. Moreover, CP and $\mathcal{R}$ parity violation makes it more complicated. Therefore, it is not viable to scan over the large parameter space of MSSM.
Fortunately, in case of generic collider searches, we do not require a very detailed descriptions of all the parameters. For example, squark/gluino pair production cross-section depends only on the squark/gluino masses and the relevant coupling. In this case, the coupling is strong coupling which is known beforehand. It is thus, possible to neglect most of the insensitive parameters and the collider phenomenology SUSY model is well described in terms of 19 parameters (p-MSSM). \\

In this article, we are mainly interested in the collider signatures of degenerate gluino-NLSP neutralino-LSP part of the SUSY parameter space. Since gluino is the NLSP, SUSY production cross-section will be dominated by gluino production. Position of the squarks in the mass spectrum also plays important role in determining the collider reach of the particular scenario. With this motivation, we consider following scenarios for our analysis. 
\begin{itemize}
\item  {\bf Scenario I: } In this scenario, we consider gluino to be NLSP and nearly degenerate with neutralino-LSP. Other sparticles are assumed to have high mass and thus, decoupled at the LHC. However, we do not consider very high mass for the squarks so that gluinos undergo prompt decay. Throughout the analysis of {\bf Scenario I}, all squarks masses are kept fixed at 3 TeV. In this case, only gluino pair production is relevant for SUSY searches at the collider. In general, the lightest neutralino is  a mixture of wino, bino and higgsino. However, the collider phenomenology does not depend on the composition of LSP. Therefore, for simplicity, we assume that LSP is purely bino like. The gluino pair production cross-section is determined by the gluino mass only. If gluino is the NLSP, gluino will undergo three-body decay to $q \bar q \tilde \chi_{1}^{0}$ via off-shell squarks. The hardness of the jets resulting from the gluino decay depends on the relative mass splitting between gluino and LSP. Therefore, the only parameters relevant for the collider phenomenology of this scenario are gluino mass and gluino neutralino mass splitting. We vary the relative splitting between LSP and gluino from 20 to 100 GeV for a particular gluino mass. In this region, the acceptance of the signal is considerably low and this is the region of our interested. The gluino decay to $q \bar{q} \chi^{0}_1$  is assumed to be 100\%. Conventional strategies for gluino search is generally applicable if the splitting is more than 100 GeV and not considered here.  
   
\item {\bf Scenario II:} This scenario is similar to {\bf Scenario I} apart from the fact that now squarks are not decoupled. Therefore, in this case, squark-gluino pair production will also contributes to the signal. Since we assume squarks to be heavier than gluino, $\tilde{q} \rightarrow \tilde{g} q$ is the dominant decay mode for the squarks. Squark-gluino pair production gives rise to five quarks at the parton level. One hard jet comes from the squark decay to gluino. Other four quarks results from the gluinos decay to nearly degenerate LSPs and thus, in most of the events, these quarks will be too soft to be detected. Therefore, squark-gluino pair production in this case gives rise to real\footnote{The jet comes from the decay of a sparticle not from the ISR/FSR.} monojet $+E_T\!\!\!\!\!/~$ signature at the collider. Both the CMS and ATLAS collaborations have already searched for the new physics signature in  monojet $+E_T\!\!\!\!\!/~$ events. This scenario could be a good new physics candidate for monojet $+E_T\!\!\!\!\!/~$ search. This motivates us to consider {\bf scenario II} and test the applicability of CMS monojet $+E_T\!\!\!\!\!/~$ search bounds for this scenario.

The inclusion of squarks makes {\bf scenario II} more complicated. Now, squark-gluino as well as gluino-gluino production cross-section also depend on the squark mass. Therefore, collider phenomenology of this scenario is determined by three parameters namely, common squark mass, gluino mass and gluino-neutralino mass splitting. However, if the mass difference between gluino and LSP is not small, it will become a conventional scenario. For this reason, the LSP mass is kept within 10 GeV below the gluino mass. We scan the parameter space over gluino mass and common squark mass. We vary common squark mass up to 2.5 TeV.  
 
\end{itemize}  

We can also think of other possibilities like degenerate squark-neutralino or degenerate 
squark-gluino-neutralino. However, the collider phenomenology of of these scenarios will be similar to the possibilities discussed above. 
One may ask about the motivation for the degenerate SUSY scenarios. Such possibilities do not appear in 
cMSSM type of scenarios. However, it usually represents non universal breaking scenario 
in which gaugino mass terms can be completely arbitrary at the high scale. There are also some well motivated 
models \cite{degenerate_gagino} in which gauginos can be nearly degenerate with each other.
Another important point is that with a  bino like LSP, it is not possible to explain current 
dark matter bound obtained from WMAP data \cite{wmap}. The neutralino is a Majorana fermion and for a 
bino like fermion, the self annihilation is done  by t channel exchange of sleptons/squarks. 
The annihilation cross section is very small and thus, a bino like neutralino LSP in general, gives rise to very high RD.
In order to reduce the dark matter density, we need some mechanism 
like resonance annihilation of dark matter particles or co-annihilation with other sparticles. 
For nearly degenerate gluino NLSP and neutralino LSP, co-annihilation is important and it 
is possible to explain current dark matter data, although the gluino and neutralino masses are always have to be fine tuned \cite{gluino_DM}.
However, it is also possible to 
satisfy relic density constraints assuming sleptons close to neutralino mass. 
As a motivation for this study, we have already discussed about another practical reason that LHC search has not found any hint of supersymmetry. As a result, it is important to know the current situation in case of degenerate mass spectrum. Since for the degenerate SUSY spectrum, the final state signature does not depend too much on the decay products of SUSY particles, the situation in this case could be drastically different from cMSSM scenario where the bounds are available from ATLAS and CMS papers. 

For {\bf Scenario I}, we consider the gluinos as the only colored particles accessible at the LHC. However, squarks can also be within the reach of the LHC. The collider phenomenology will be different in the case where squark-gluino production also contributes to the signal. Therefore, it is also important to understand the gluino mass reach in presence of squarks. In {\bf Scenario II}, we consider such a situation in which gluino is NLSP and nearly degenerate with the LSP as in the case of {\bf Scenario I}. Moreover, we consider squarks which are heavier than gluino however, accessible at the LHC.\\

\section{Analysis}

Before going into the details of our analysis, let us briefly introduce the monojet \cite{monojet_CMS} and multijet search \cite{multijet_ATLAS} strategies and results of CMS and ATLAS collaboration respectively. 
\begin{itemize}
\item {\bf CMS monojet search} \cite{monojet_CMS}: CMS collaboration have performed a search for an energetic jet $+ E_T\!\!\!\!\!\!/~$ at the LHC with center-of-mass energy 7 TeV and 5.0 inverse femtobern integrated luminosity. Events with at least one jet ($j_1$) with $p_T>110$ GeV and $E_T\!\!\!\!\!\!/~>200$ GeV are selected for the further analysis. 
Dijet events are allowed if the angular separation between two jets satisfies $\Delta\phi(j_1 , j_2 ) < 2.5$ radian. Events with a third jets with $p_T>30$ GeV are discarded. To reduce the background from $W$ and $Z$ production, events with isolated electrons, muons and charged tracks with $p_T>10$ GeV 
are also rejected. With these event selection criteria, the number of event observed by the CMS collaboration is consistent with the number of events expected from the SM in the all $E_T\!\!\!\!\!\!/~$ range. In view of this consistency, 95\% C.L. upper limits for different $E_T\!\!\!\!\!\!/~$ cuts are set on the number of events arising from the non-SM processes. The upper limits on the number of non-SM events are presented in Table~\ref{CMS_limits}.

\begin{table}[h]

\begin{center}

\begin{tabular}{||c||c|c|c|c||}
\hline \hline
$E_T\!\!\!\!\!\!/~$ in GeV & $\ge 250$ & $\ge 300$  & $\ge 350$  & $\ge 400$\\\hline\hline
Expected limit & 779  &  325  &  200  &  118 \\
Observed limit & 600  &   368 &    158&     95 \\\hline\hline
\end{tabular}

\end{center}

\caption{Expected and observed upper limits set by the CMS collaboration \cite{monojet_CMS} on the number of non-SM events passing the event selection criteria and $E_T\!\!\!\!\!\!/~$ cuts. The numbers correspond to the LHC with $\sqrt s=7$ TeV and integrated luminosity 5.0 fb$^{-1}$.}

\label{CMS_limits}
\end{table}

\item {\bf ATLAS multijet search} \cite{multijet_ATLAS}: A search for 2-6 jets in association with large $E_T\!\!\!\!\!\!/~$ at the LHC with $\sqrt s=7$ TeV and 4.7 inverse femtobern integrated luminosity has been communicated by the ATLAS collaboration. Jet candidates are reconstructed using the anti-kt jet clustering algorithm~\cite{anitkt} with a distance parameter of 0.4 in the rapidity coverage $|\eta|\le 4.9$. Electron (muon) candidates are required to have $p_T > 20(10)$ GeV and $|\eta| < 2.47(2.4)$. After identifying jets and lepton, any jet candidate lying within a distance $\Delta R = \sqrt{\Delta \eta^2 + \Delta \phi^2} < 0.2$ of an electron is discarded. A lepton candidate is removed from the list is it falls with in a distance $\Delta R=0.4$ of any survived jet candidate. Missing transverse momentum is reconstructed using all remaining jets and leptons and all calorimeter clusters not associated to such objects. Finally, jets with $|\eta|>2.8$ are removed from the list. After the object reconstruction, events with zero electron (muon) with $p_T>20(10)$ are selected for further analysis.

ATLAS collaboration has presented results for five inclusive analysis channels, characterized by increasing jet multiplicity from 2 to 6. In our analysis, we only consider dijet and trijet channels. In Table~\ref{ATLAS_limits}, we have presented the cuts used by the ATLAS collaboration to define the signal regions. For all the signal regions (SRs) defined by the ATLAS collaboration, good agreement is seen between the numbers of events observed in the data and the numbers of events expected from SM processes. As a result, 95\% C.L. upper limits are set on the  beyond SM cross-sections ($\sigma_{BSM}$) for different SRs.


\begin{table}[h]

\begin{center}

\begin{tabular}{||c||c|c|c|c|c|c|c||c|c||}
\hline \hline
 & \multicolumn{7}{|c||}{cuts} & \multicolumn{2}{|c||}{Upper limits}\\\cline{2-8}
Channel & $E_T\!\!\!\!\!\!/~$ & $p_T^{j_1}$ & $p_T^{j_2}$ & $p_T^{j_3}$ & $\Delta \phi$ & ${E_T\!\!\!\!\!\!/~}/$ & {$m_{eff}^{incl.}$} & \multicolumn{2}{|c||}{on $\sigma_{BSM}$ [fb]}\\\cline{9-10}
        & GeV & GeV & GeV & GeV &$(jet,E_T\!\!\!\!\!\!/~)$ & ${m_{eff}(N_j)}$ & {GeV} & Exp. & Obs. \\\hline\hline
$A$ tight  & & && - &  & 0.3 (2j) & 1900  & 1.3 & 0.62 \\
$A$ medium & & && - &  & 0.3 (2j) & 1400  & 6.0 & 5.3 \\
$A^\prime$ medium & 160 & 130 & 60 & - & 0.4  & 0.4 (2j) & 1200 & 9.2 & 6.2 \\
$B$ tight &   & && 60 &  & 0.25 (3j) & 1900 &  1.2 & 0.65 \\\hline\hline

\end{tabular}

\end{center}

\caption{Cuts used by the ATLAS collaboration to define the signal regions for dijets and trijets channel. $\Delta \phi(jet,E_T\!\!\!\!\!\!/~)$ is the azimuthal separations between $\vec p_T\!\!\!\!\!/~$ and the reconstructed jets. ${m_{eff}(N_j)}$ is defined to be the scalar sum of the transverse momenta of the leading $N$ jets together with $E_T\!\!\!\!\!\!/~$. However, for $m_{eff}^{incl.}$, the sum goes over all jets with $p_T>40$ GeV. Last two column corresponds to the 95\% C.L. expected (Exp.) and observed (Obs.) upper limits on the non-SM contributions.}

\label{ATLAS_limits}
\end{table}

 
\end{itemize}

We are considering the scenarios in which hard processes gives rise to small number of jets in the final state. Therefore, we only concentrate on the applicability of CMS monojet $+E_T\!\!\!\!\!\!/~$ and ATLAS multijet (2 and 3-jets only) $+E_T\!\!\!\!\!\!/~$ search bounds for our scenarios. 

For scenario I, we consider gluino to be nearly degenerate with neutralino and squarks are too heavy. Therefore, gluino pairs will be copiously produced at the LHC. However, due to small gluino-neutralino mass splitting, gluino decays to neutralino gives rise to jets which are too soft to pass the event selection criteria of CMS and ATLAS collaboration. Therefore, for scenario I, we have generated $\tilde g \tilde g$ and $\tilde g \tilde g$ in association with one jet\footnote{Since for the ATLAS multijet analysis, we are considering 2- and 3-jet final state, we should generate gluino pairs up to 2/3 additional jets at the parton level. However, we have checked that the result does not vary significantly if we consider only one additional jet instead of up to 2/3 jets at the parton level.} events at the parton level using {\bf MadGraph} 5 \cite{madgraph}. Subsequently, Madgraph generated events are passed into {\bf PYTHIA} \cite{pythia} for matching using MLM prescription \cite{MLM}, simulating the decay, hadronization e.t.c. We have 
used {\bf Prospino} 2.1 \cite{prospino} with CTEQ6.6M \cite{cteq6.6m} parton distribution function (pdf) for computing NLO gluino pair production cross-section. The QCD factorization and renormalization scales are kept fixed at twice gluino mass.

In scenario II, we investigate the situation where gluino is still nearly degenerate with neutralino, however, squarks are not too heavy. In this case, squark-gluino and squark-squark production also contribute to the signal. Moreover, unlike in the case of gluino pair production, in this case squark-gluino and squark-squark production give rise to one or two hard jets arising from squarks decay to gluino. For simulating gluino pair production, we have used the same prescription discussed the previous paragraph. For squark-gluino and squark-squark production, hard jets are produced from squarks decay to gluino. Therefore, we use {\bf PYTHIA} for generating parton level squark-gluino and squark-squark events as well as for simulating the squark and gluino decay, ISR, FSR, hadronization e.t.c. Here also we have used {\bf Prospino} 2.1 for calculating NLO production cross-sections for $\tilde g \tilde g$,  $\tilde q \tilde g$, $\tilde q \tilde q$, $\tilde q \tilde q^*$. The QCD factorization and renormalization scales are kept fixed at the sum of masses of the produced SUSY particles. For scenario II, we only consider the production of the first two generation of squarks.

After showering and hadronization in {\bf PYTHIA}, the events are fed to fast 
detector simulator package {\bf Delphes} \cite{delphes} for object (jets, leptons, $E_T\!\!\!\!\!\!/~$  e.t.c.) reconstruction. We have closely followed CMS (for monojet analysis) and ATLAS (for multijet analysis) collaborations suggested  object reconstruction criteria and cuts. The brief description of our object reconstruction is as follows:
\begin{itemize}
\item {\bf Jets:} For CMS monojet (ATLAS multijets) bound, jets are constructed using anti-kt algorithm with R parameter to be equal to 0.5 (0.4) and only jets with $p_T>$ 30 (20) GeV and $|\eta|<$ 5.0 (4.9) are considered for further analysis.
\item {\bf Leptons:} We demand that lepton candidates (both electron and muon) are required to have $p_T > 20$ GeV/c and to be separated from jets by at least 
$\Delta R=0.5$ for monojet analysis. For multijets analysis, we have used  ATLAS definition for leptons and lepton-jet isolation discussed in the previous paragraph.
\item {\bf Missing $E_T$:}   The missing $E_T$ in an event is calculated using calorimeter cell energy and the momentum of the reconstructed muons in the muon spectrometer. 
\end{itemize}

The dominant SM backgrounds for monojet as well as multijet plus missing energy 
search arises from  $Z(W)$ + jets production followed by $Z$ decay to $\nu \bar{\nu}$ and  $W \rightarrow l \nu $. The first one is the real background to the jets plus missing energy (plus 0 lepton ) signature whereas, the second one contributes only when the lepton is not reconstructed. The effect of top quark is negligible for monojet search because of higher jet multiplicity from the decay of top quark. For multijet search vector boson pair production and $t\bar{t}$ is not negligible. In both cases QCD multijet background is reduced to negligibly small value by applying high $E_T\!\!\!\!\!\!/~$ and effective mass ($m_{eff}$) cuts.  The SM backgrounds events ($Z$ + jets, $W$ + jets, $t\bar{t}$  and vector boson pair productions) are generated by {\bf MadGraph} 5. Subsequently, the {\bf MadGraph} generated SM background events are analyzed by {\bf PYTHIA} and {\bf Delphes}.
\section{Result and discussion}

After introducing the CMS and ATLAS monojet and multijet $+E_T\!\!\!\!\!\!/~$ search strategies as well as our signal and background analysis methods, we are now equipped enough to discuss the final out come of our analysis. To constrain the parameter space of {\bf Scenario I} and {\bf Scenario II} from the 7 TeV LHC data, we use 95\% C.L. upper limits set by the CMS and the ATLAS on beyond SM contributions ($\sigma_{BSM}$) to the monojet $+E_T\!\!\!\!\!\!/~$ (listed in Table~\ref{CMS_limits}) and multijet $+E_T\!\!\!\!\!\!/~$ (listed in Table~\ref{ATLAS_limits}) final states respectively. Since the jet multiplicity is expected to be small for the SUSY scenarios under consideration, we only consider the first four signal regions (namely A-Tight, A-Medium, A$^{\prime}$-Medium and B-Tight) of ATLAS analysis. 

We have also calculated the expected 99.7\% C.L. ($3\sigma$) discovery reach for these scenarios at the LHC with $\sqrt s=8$ TeV and 15 fb$^{-1}$ integrated luminosity. For the multijets $+E_T\!\!\!\!\!\!/~$ analysis, we have used the ATLAS event selection criteria discussed in the previous section and in Table~\ref{ATLAS_limits}. However, for the monojet $+E_T\!\!\!\!\!\!/~$ analysis, we have used the CMS cuts in Table~\ref{CMS_limits} as well as we have added two more signal regions defined by $E_T\!\!\!\!\!\!/~\ge 500,~{\rm and}~600$ GeV. In this analysis, we consider the SUSY signature to be observable over the SM background at 99.7\% C.L. if $\sigma_{SUSY}/\sqrt{\sigma_{SM}} \ge 3/\sqrt{\cal L}$, where $\sigma_{SUSY}$ and $\sigma_{SM}$ are the SUSY and the SM contributions to the respective final states and  ${\cal L}$ is the integrated luminosity. \\
\begin{figure}[h]
\begin{center}
\epsfig{file=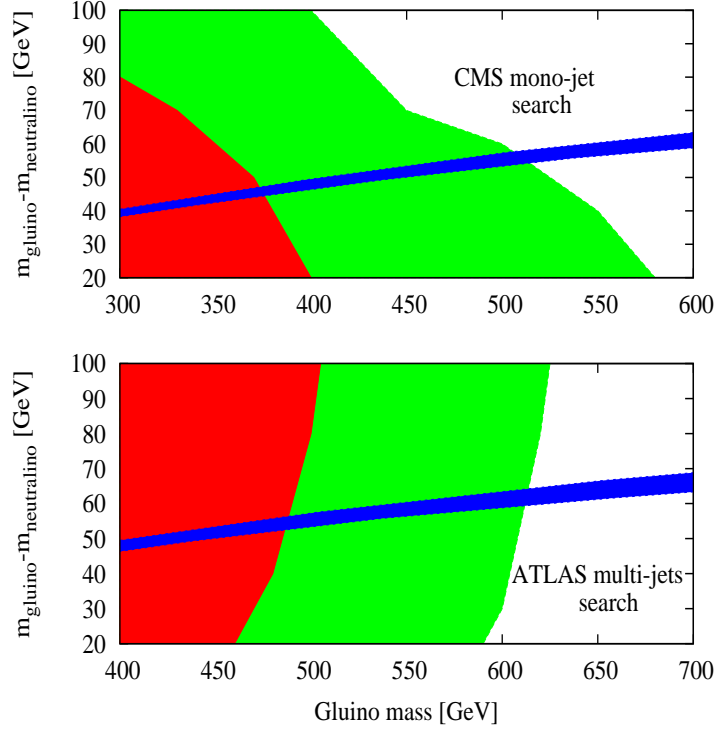,width=10cm,height=10cm,angle=270}
\end{center}
\caption{For {\bf Scenario I}, 95\% C.L. observed exclusion limit (red region) from 7 TeV LHC data and 99.7\% C.L. expected reach (green region) at the LHC with center-of-mass energy of 8 TeV and 15 fb$^{-1}$  integrated luminosity. We have considered both the CMS and the ATLAS monojet (top panel) and multijet (bottom panel) $+E_T\!\!\!\!\!\!/~$ search bounds from 7 TeV LHC data with integrated luminosity of 5.0 and 4.7 fb$^{-1}$ respectively.}
\label{fig:GG_ATLAS}
\end{figure}

\noindent {\bf Scenario I:} In this scenario, gluino is NLSP and nearly degenerate with the neutralino-LSP. Other sparticles are decoupled at the LHC. Therefore, the collider phenomenology of this scenario is determined by two parameters namely, the gluino mass and gluino-neutralino mass splitting. Since we consider ${\cal R}$-parity conservation, the neutralino-LSP contributes to the dark matter density. The dark matter phenomenology depends on the composition (wino, bino and higgsino component) of the LSP. In this analysis, we assume the LSP to be purely bino like. We present a two dimensional scan over gluino mass ($m_{\tilde g}$) and gluino-neutralino mass splitting ($\Delta m_{\tilde g-\tilde \chi_1^0}=m_{\tilde g}-m_{\tilde \chi_1^0}$) in Fig.~\ref{fig:GG_ATLAS} which will be discussed in details in the following.
\begin{itemize}
\item {\bf Fig.~\ref{fig:GG_ATLAS} (top panel)}: The red region corresponds to the part of $m_{\tilde g}-\Delta m_{\tilde g-\tilde \chi_1^0}$ plane excluded at 95\% C.L. from the CMS monojet $+E_T\!\!\!\!\!\!/~$ search at 7 TeV center-of-mass energy and 5.0 fb$^{-1}$ integrated luminosity. The green region corresponds to the 99.7\% C.L. discovery reach of the LHC running at $\sqrt s=8$ TeV and 15 fb$^{-1}$ integrated luminosity. For the CMS monojet $+E_T\!\!\!\!\!\!/~$ analysis, events with more than two jets are not considered for the analysis. Increasing gluino-neutralino mass splitting implies harder jets arising from the gluino decay and thus, increasing final jet multiplicity. As a result, for CMS  monojet $+E_T\!\!\!\!\!\!/~$ search, gluino mass reach decreases as we increase the gluino-neutralino mass splitting.

\item {\bf Fig.~\ref{fig:GG_ATLAS} (bottom panel)}: In this figure, we consider the ATLAS multijets $+E_T\!\!\!\!\!\!/~$ search channels. The excluded part of the parameter space from 7 TeV run and expected discovery reach at the 8 TeV LHC with 15 fb$^{-1}$ integrated luminosity are indicated by the red and green regions. Larger $\Delta m_{\tilde g-\tilde \chi_1^0}$ corresponds to the harder jets from gluino decay and thus increasing possibility of surviving ATLAS multijets $+E_T\!\!\!\!\!\!/~$ selection criteria. In Fig.~\ref{fig:GG_ATLAS} (bottom panel), the gluino mass bound/reach increases slightly for large gluino-neutralino mass splitting.

\item In Fig.~\ref{fig:GG_ATLAS}, the blue band corresponds to the region consistent with the WMAP \cite{wmap} relic density data. With the combined data from WMAP, BAO (Baryon Acoustic Oscillations) in the distribution of galaxies and observation of Hubble constant, the density of cold dark matter in the
universe is determined to be $\Omega_ch^2=0.1126\pm0.0036$. We consider the preferred RD range of $0.0941<\Omega_ch^2<0.1311$ at $2\sigma$ level. We have used {\bf micrOmegas} (v.2.4.R) \cite{microomega} for calculating RD in this scenario.
\end{itemize}

\begin{figure}[h]
\begin{center}
\epsfig{file=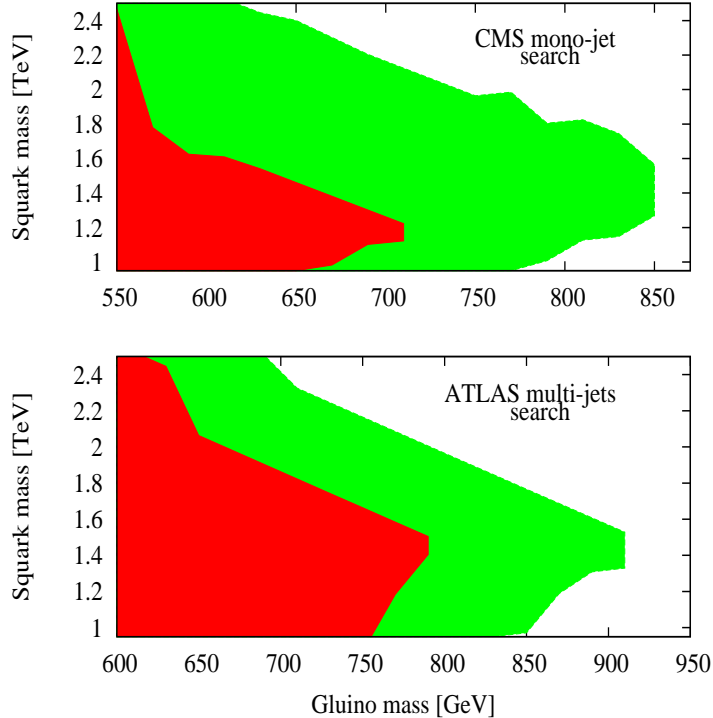,width=10cm,height=10cm,angle=270}
\end{center}
\caption{For {\bf Scenario II}, 95\% C.L. observed exclusion bound from 7 TeV LHC data and 99.7\% C.L. expected exclusion bound on the squark mass as a function of gluino mass at the proton-proton collision with center-of-mass energy of 8 TeV and 15 fb$^{-1}$  integrated luminosity. We have considered both the CMS and the ATLAS monojet (top panel) and multijet (bottom panel) $+E_T\!\!\!\!\!\!/~$ search bounds at the center-of-mass energy of 7 TeV and integrated luminosity of 5.0 and 4.7 fb$^{-1}$ respectively.}
\label{fig:SG_ATLAS}
\end{figure}

\noindent {\bf Scenario II:} For this scenario, we consider the squarks to be heavier than the gluino however, not too heavy to be produced at the LHC. Here also gluino is the NLSP and nearly degenerate with the neutralino-LSP. Therefore, the collider phenomenology in this case, depends on three parameters namely, squark mass, gluino mass and gluino-neutralino mass splitting. Fig.~\ref{fig:GG_ATLAS} (top panel) shows that gluino-neutralino mass splitting has important impact for the CMS monojet $+E_T\!\!\!\!\!\!/~$ search for {\bf Scenario I}. However, it is important to note that at the parton level, the pair production of gluinos in {\bf Scenario I} does not contribute to the monojet $+E_T\!\!\!\!\!\!/~$ final state. Monojet signature in {\bf Scenario I} arises when few jets from the gluinos decay fall outside detector coverage and a hard jet results from the ISR/FSR. If the gluino is nearly degenerate with the LSP, in most of the events, the jets from the gluino decay will be too soft to be detected. However, squark-gluino production in {\bf Scenario II}, contributes to the monojet $+E_T\!\!\!\!\!\!/~$ final state without a hard jet from the ISR/FSR. In this case, one hard jet results from the squark decay to the gluino. Fig.~\ref{fig:GG_ATLAS} (bottom panel) shows that gluino-neutralino mass splitting does not play significant role for the ATLAS multijets $+E_T\!\!\!\!\!\!/~$ searches as long as the gluino-neutralino mass splitting is not significantly large. Therefore, in this part of our analysis, we fix the gluino-neutralino mass splitting at 10 GeV and scan over gluino and squark masses. The result is presented in Fig.~\ref{fig:SG_ATLAS} and discussed in the following.

\begin{itemize} 
\item {\bf Fig.~\ref{fig:SG_ATLAS} (top panel)}: We have presented the region (red region) in the squark-gluino mass plane excluded at 95\% C.L. from CMS monojet $+E_T\!\!\!\!\!\!/~$ search at $\sqrt s=7$ TeV and 5.0 fb$^{-1}$ integrated luminosity data. The green region corresponds to the expected 99.7\% C.L. reach in the same channel at the LHC with 8 TeV center-of-mass energy and 15 fb$^{-1}$ integrated luminosity. 
\item {\bf Fig.~\ref{fig:SG_ATLAS} (bottom panel)}: Same as above but for the ATLAS multijets $+E_T\!\!\!\!\!\!/~$ search channels. 
\item Fig.~\ref{fig:SG_ATLAS} shows that for both  the CMS monojet and  the ATLAS multijets $+E_T\!\!\!\!\!\!/~$ search channels, squark mass bound/reach decreases sharply as we increase the gluino mass in the low gluino mass region. In the low gluino mass and high squark mass region, dominant contributions to the signal arises from the gluino pair production. In this part, squark mass reach falls rapidly with the gluino mass. However, for intermediate squark masses, squark-gluino production significantly contributes to the signal. In this part of the parameter space, squark-gluino pairs are produced with a non-negligible rate. Moreover, the signal acceptance efficiency is high in this region because a hard jet arises from squark decay to gluino.
\item For low squark masses, the squark-gluino production cross-section enhances however, the signal acceptance efficiency decreases. As a result, Fig.~\ref{fig:SG_ATLAS} shows that for high gluino mass, low squark mass regions can not be excluded/probed with present CMS or ATLAS monojet or multijet
 $+E_T\!\!\!\!\!\!/~$ search approaches.  

\end{itemize}

After discussing the phenomenology of {\bf Scenario I} and {\bf II} as well as describing Fig.~\ref{fig:GG_ATLAS} and \ref{fig:SG_ATLAS}, we can now summarize our result. For {\bf Scenario I}, Fig.~\ref{fig:GG_ATLAS} shows that for low gluino-neutralino mass splitting, gluino mass below about 400 (470) GeV is excluded at 95\% C.L. from 7 TeV CMS (ATLAS) monojet (multijets) $+E_T\!\!\!\!\!\!/~$ search. Whereas, for high gluino-neutralino mass splitting ($\sim ~80$ GeV), the gluino mass is excluded up to only 300 GeV from the CMS monojet search 500 GeV from ATLAS multijets search. Fig.~\ref{fig:GG_ATLAS} also shows that the LHC with $\sqrt s=8$ TeV and 15 fb$^{-1}$ integrated luminosity will be able to extend the gluino mass reach by about 150 GeV only. For {\bf Scenario II}, Fig.~\ref{fig:SG_ATLAS} shows that for gluino mass of 650 (750) GeV, squark masses up to 1.5 (1.6) TeV is excluded. At this point, we would like to draw the attention towards the following points:
\begin{itemize}
\item Both Fig.~\ref{fig:GG_ATLAS} and \ref{fig:SG_ATLAS} shows that multijets $+E_T\!\!\!\!\!\!/~$ search channels have better reach than the monojet $+E_T\!\!\!\!\!\!/~$ search channels for both the scenarios. The CMS monojet $+E_T\!\!\!\!\!\!/~$ analysis does not consider events with more than two jets. We find that this criteria kills the SUSY signal significantly because there will be always a few hard jets arising from the ISR/FSR as well 
as from the decay of gluino/squarks.
 
\item To reduce the SM background, the ATLAS collaboration have used very high effective mass cuts for their multijets $+E_T\!\!\!\!\!\!/~$ analysis. The ATLAS cuts are optimized for the cMSSM scenario where the mass splittings between squarks/gluino and LSP are in general high and thus cMSSM scenario gives rise to very hard effective mass distribution. However, degenerate SUSY scenarios mostly gives rise to few soft jets and low $E_T\!\!\!\!\!\!/~$ and thus low effective mass. We find that ATLAS collaboration used high effective mass cuts severely kill the degenerate SUSY signal. As a result, the ATLAS reach also is not very good for the degenerate scenarios (see Fig.~\ref{fig:GG_ATLAS} and \ref{fig:SG_ATLAS}). We also notice that the situation does not improve much at the LHC with $\sqrt s=8$ TeV. We also check that moderate change in the effective mass cut and $E_T\!\!\!\!\!\!/~$ do not 
improve the reach significantly. 

\item  We can see a clear discrepancy between Fig.~\ref{fig:GG_ATLAS} and \ref{fig:SG_ATLAS}. Fig.~\ref{fig:SG_ATLAS} should reproduce the gluino bound as in Fig.~\ref{fig:GG_ATLAS} in the limit of very heavy squark mass. However Fig.~\ref{fig:SG_ATLAS} shows somewhat larger value of gluino
mass in that case. For example, Fig 2  shows that 
the gluino mass bound from 7 TeV data is about 600 GeV whereas the
bound in Fig 1 is about 500 GeV. The reason is the following.
In Fig 1 we use matching in the gluino production and in Fig 2 there is
no matching. Thus in the decoupled squark limit the gluino mass 
bound has large uncertainty in the later case. We have checked that 
the bound is strongly dependent on the choice of factorization scale 
and thus not reliable. However, the parameter space with moderate value of
squark mass can be used safely.

\item Besides supersymmetry there exist BSM model like Universal Extra Dimension
which can naturally predict degenerate spectrum. In the minimal version
of UED, the overall splitting is determined by two parameters : radius of
compactification $R^{-1}$ and cut-off scale $\Lambda$. For moderate
splitting ($\Lambda R \sim$  20 ) the leptonic channels are efficient
to probe such models \cite{Bhattacherjee:2010vm,Murayama:2011hj}. However recent study shows that in order to explain
electroweak data as well as 125 GeV Higgs boson mass we need very small
value of $\Lambda R$ ($\sim$ 2-3) \cite{Datta:2012db}. In this case the leptons that come
from the decay of $n=1$ KK particles will be very soft and may not be
detected. Even in that case ISR radiations may be hard and it is
possible to discover or exclude UED model using monojet and multijet
search. Our naive estimate shows that we can exclude $R^{-1}$ up to
600 GeV of minimal UED model for all values of $\Lambda R$. However
again in that case matching will play crucial role. A detailed study
is required in this case.      
\end{itemize}

\section{Conclusion}We have investigated the collider phenomenology of two degenerate SUSY scenarios: (i) {\bf Scenario I}: gluino is the NLSP and nearly degenerate with the LSP and all other sparticles are assumed to be too heavy to be produced at the LHC and (ii) {\bf Scenario II}: which is same as {\bf Scenario I} but squarks are considered to be within the reach of the LHC. In the context of these degenerate SUSY scenarios, we have studied the impact of $\sqrt s=7$ TeV and 5.0 (4.7) fb$^{-1}$ integrated luminosity monojet (multijets) $+E_T\!\!\!\!\!\!/~$ search bounds from CMS (ATLAS) collaborations. We have also investigated the discovery reach of these scenarios at the LHC with $\sqrt s=8$ TeV.

We have found the the CMS and the ATLAS jets $+E_T\!\!\!\!\!\!/~$ search strategies are not optimized for the degenerate 
beyond SM scenarios and the discovery reach for these scenarios do not improve much at the LHC with $\sqrt s=8$ TeV if 
we stick to the conventional techniques. Cut optimization may improve the situation and more detailed study is required 
in this case which is beyond our scope. Therefore, we request our experimental colleagues to study degenerate BSM models 
and improvise the analysis technique although it could be very challenging.

\section{Note added} 
While this paper was in preparation, we came across the reference \cite{Dreiner}, where the 7 TeV LHC bounds on nearly degenerate gluino NLSP and neutralino LSP scenario have been studied. Therefore, small part of our result (7 TeV bounds on the parameters of {\bf Scenario I}) has overlap with Ref.~\cite{Dreiner}. It provides an independent confirmation of some of our result. However, in this paper, we have also 
investigated the collider phenomenology of {\bf Scenario II}. We have investigated the expected reach at the LHC with $\sqrt s=8$ TeV.

\vskip 20pt

{\bf Acknowledgments:} \\
The work of BB is supported by World Premier International Research Center Initiative (WPI
Initiative), MEXT, Japan. KG acknowledge support from the Academy of Finland (Project No. 137960). The authors would like to thank the organizers of WHEPP12 (Mahabaleswar, India, 2012), and University of Calcutta, India, for the hospitality at the 
initial part of this work. BB would like to thank Prof. M Nojiri, Prof. S. Matsumoto, K Harigaya 
for useful discussions on Madgraph and Delphes software.

\end{document}